\documentclass[twocolumn,twoside,letterpaper,11pt]{article}
\usepackage{graphicx,xrrc,natbib}
\usepackage{times}

\begin{document}


\title{DYING RADIO GALAXIES IN CLUSTERS}


%
%
%
%


\author{   M. Murgia                        } 
\institute{   Istituto di Radioastronomia del CNR}
\institute {Osservatorio Astronomico di Cagliari, INAF       } 
\address{     Via Gobetti 101, I-40129 Bologna, Italy }
\address{ Loc. Poggio dei Pini, Strada 54, I-09012 Capoterra (CA), Italy}

\email{    murgia@ira.cnr.it                        } 

\author{    P. Parma, H.R. de Ruiter, K.-H. Mack, R. Fanti }
\email{     parma@ira.cnr.it, deruiter@ira.cnr.it, mack@ira.cnr.it, rfanti@ira.cnr.it}


\maketitle

\abstract{  We report the recent discovery of three `dying' radio galaxies belonging to the
WENSS minisurvey sample: WNB1734+6407,  WNB1829+6911 and WNB1851+5707. These
sources have been selected on the basis of their extremely steep broad-band radio spectra, which
 is a strong indication that these objects either belong to the rare class
of dying radio galaxies or that we are observing `fossil' radio plasma
 remaining from a previous nuclear activity. Deep spectral index images obtained with the
Very Large Array confirmed that in these sources the central engine has ceased to be active for
 a significant fraction of their lifetime although their extended lobes have
not yet completely faded away. In one case, WNB1829+6911, fossil radio lobes are seen in conjunction
 with newly restarting jets.  We found that each source is located (at least in projection) at the
center of an X-ray emitting cluster. We argue that their intriguing association with
clusters implies that the pressure of the dense intracluster medium, perhaps a
cooling flow, prevents quick liquidation of a fossil radio lobe through
adiabatic expansion. On statistical ground we deduce that the duration of the dying phase for a radio source in
cluster is one order of magnitude higher with respect to that of a radio galaxy in the field.}

\section{Introduction}

\begin{table*}[t]
  \begin{center}
  \caption{Dying radio galaxies and their hosting clusters. Spectral indexes and flux densities refer to integrated values.}\medskip
  \label{tab:pages}
  \begin{tabular}{ccccccccc}
    WENSS name & ID &Cluster & redshift & $S_{1.4 \rm GHz}$  & $\alpha$ & LAS&LS & L$_{151 \rm MHz}$\\
                                                                                                                &     &             &              &    (mJy)                  &     (at 1.4 GHz)         &(arcsec)& (kpc)& W\,Hz$^{-1}$\\
    \hline\hline
 WNB1734+6407   & G  &A2276              & 0.1406  & 7.2   &  2   & 50  & 122& $10^{25.4}$\\
 WNB1829+6911   & G  &ZwCl 1829.3+6912   & 0.204   & 12.6  &  1.9 & 40  & 133& $10^{25.9}$\\
 WNB1851+5707   & G  &RXC J1852.1+5711   & 0.109   & 53.3  &  1.7 & 20  & 39 & $10^{25.3}$\\

    \hline
  \end{tabular}
  \end{center}
\end{table*}

Strong radio sources associated with elliptical galaxies are supplied with
energy from active galactic nuclei via plasma beams. If this energy supply
ceases, a source is expected to undergo a period of fading before it
disappears completely. In this phase, sources may lack certain features, such as
radio cores or well-defined jets that are commonly interpreted as indicators
of continuing activity. Alternatively, it is possible that radio galaxies
may be active intermittently.  In this scenario, one expects to observe
fossil radio plasma remaining from an earlier active epoch, along with newly
restarting jets. The best case for fossil radio lobes seen with a currently active galaxy
is 3C 338 (Jones \& Preston 2001). The very steep spectrum lobes of this source
 are clearly disconnected from the currently active jets.
In both scenarios, the fading lobes are expected to have
very steep ($\alpha > 1.5$, $S_{\rm \nu} \propto\nu^{-\alpha}$) and convex
radio spectra characteristic of a population of electrons which have
radiated away much of their original energy (Komissarov \& Gubanov 1994).
Only a handful of fossil radio galaxies with the
characteristics described above are known  (e.g. Cordey 1987, Venturi et al. 1998).
In their sample of radio sources selected from the B2 and 3C catalogs
 Giovannini et al. (1988) found that only a few percent of objects
 exhibit the characteristics mentioned above.
 A possible explanation for the rarity of the fading radio galaxies may be the short
duration of the remnant phase of a radio source with respect to the average
lifetime of the radio activity. Synchrotron losses and
the inverse Compton scattering of the Cosmic Microwave Background photons
preferentially deplete the high-energy electrons. In the absence of fresh
particle injection, the high-frequency radio spectrum develops an
exponential cutoff. At this point, the adiabatic expansion of the
 radio lobes will concur to shift this spectral break to lower frequencies
 and the source will disappear quickly. On the other hand, if the source expansion
is somehow reduced, or even stopped, there is still the chance to detect the
fossil radio lobe, at least at low frequency. It is important to note that these sources
represent the last phase in the `life' of a radio galaxy. Therefore they must be well distinguished from
the cluster radio halo/relic phenomenon, that is usually not associated with an
 individual galaxy (Slee et al. 2001).

\section{Source selection and integrated spectra}
For the reasons mentioned above, the Westerbork Northern Sky Survey (WENSS) at 327 MHz is particularly
well-suited to search for these elusive fossil radio sources.  A sub-sample, the WENSS
minisurvey (de Ruiter et al.\ 1998), contains 400 radio sources identified
with either elliptical or spiral galaxies with red magnitude brighter than
$m_{\rm r}= 16.5$. Using cross-correlation with the existing catalogues and
new observations with the Effelsberg 100-m telescope, we have obtained spectral information in the frequency
range 38 MHz to 10 GHz for 200 bright sources.  Three out of 200 sources
show extremely steep spectra which are characterized by a quasi-exponential drop
above a frequency of about 1 GHz. These are WNB1734+6407, WNB1829+6911,
WNB1851+5707  (see Tab. 1 and Fig.~1).

\begin{figure*}[t]
  \begin{center}
    \includegraphics[width=17cm]{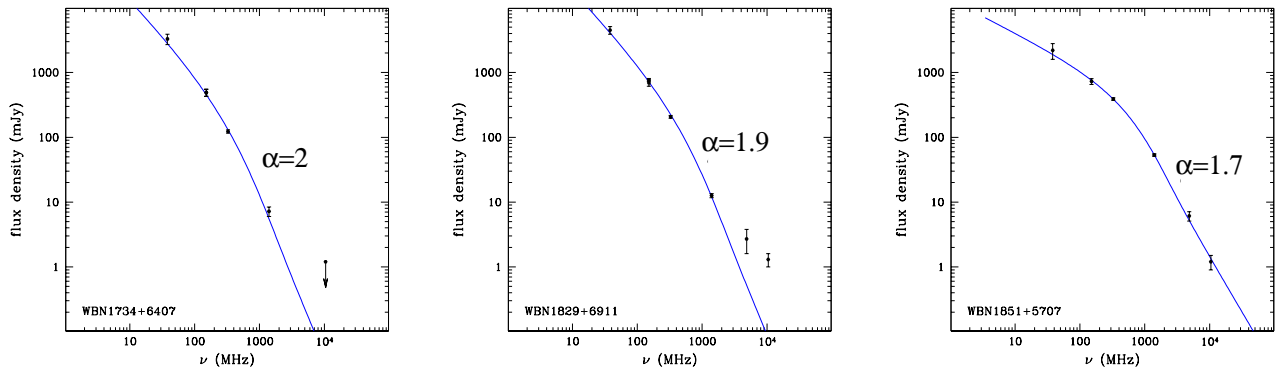}
    \caption{\small  The ultra-steep spectra of WNB1734+6407, WNB1829+6911 and
  WNB1851+5707. Compare with the typical active radio galaxies which have
  power law spectra with $\alpha \simeq 0.8 $.  The solid lines represent the best fit of the
Komissarov-Gubanov model (see text). The relative durations of the relic phase ($t_{\rm RE}$) with
respect to the active phase ($t_{\rm CI}$) are in the range $0.5-0.7$.}

    \label{fig1}
  \end{center}
\end{figure*}

We have fitted  the spectral aging model of Komissarov \& Gubanov (1994)
to the  extremely steep broad-band radio spectra of the three sources.
They assumed a finite duration for the continuous
 injection (CI) of relativistic electrons, $t_{\rm CI}$, followed by a relic
 phase (RE) of duration $t_{\rm RE}$. During both the CI and the RE phase,
 the electrons lose energy by synchrotron emission and inverse Compton scattering of
cosmic microwave background photons. The fit of the Komissarov-Gubanov model to the radio
spectra support  the scenario in which the injection of fresh electrons in these sources has
ceased for a significant fraction of their lifetime ($t_{\rm RE}/t_{\rm CI} \sim 0.5 - 0.7$).

\section{Very Large Array spectral index imaging}
In order to determine whether these sources were really dying objects or
relic lobes associated with active radio galaxies,
we observed their radio continuum emission at 20 and 6 cm with the Very Large Array (VLA) in various configurations.
These extensive campaign of observations allowed us to obtain very deep spectral images of matched angular
 resolution, see right panels of Fig. 2.  We achieved a noise level of about 10 $\mu{\rm Jy/beam}$ and an angular resolution of 6 arcsec
 at both frequencies.
For all three sources, the deep VLA images confirmed beyond any doubt the presence
of fossil radio plasma from an earlier nuclear activity. It is worth noting that all
 these radio sources have a firm optical identification with an elliptical galaxy, but, at the
 sensitivity limit of our observation, none of them possesses radio jets on kpc scale.

Characterized by two relaxed lobes lacking hot-spots, the radio morphology of WNB1734+6407
resembles that of B2 0924+32 which is considered the prototype of fossil radio galaxies (Cordey 1987).
The radio spectra of the fossil lobes of WNB1734+6407 is so steep that their surface brightness at
 6 cm is below the sensitivity level of our observations. Thus, we can only place lower limits on
  the spectral index which result in $\alpha>3.3$ and  $\alpha>2.1$ for the northern and southern lobe,
        respectively. We also observe a slightly extended component coincident with the galaxy core. Also this
        feature has a quite steep spectral index, $\alpha=1.8$, and its nature remains unclear.

The radio appearance of WNB1829+6911 is virtually identical to that of 3C338, a
nearby radio source associated with the central dominant galaxy in the cooling flow
cluster Abell 2199. In both sources we observe the presence of fossil plasma remaining
 from a previous activity in conjunction with a restarting core.  The extended emission of
  WNB1829+6911 has a very steep spectrum, $\alpha > 2.2$. The source exhibits a bright core with a
        much flatter spectrum, $\alpha = 0.6$. Most probably this region of the radio source is powered
        by a new couple of restarting jets. The core emission is responsible for the high-frequency flattening
        seen in the source integrated spectrum (Fig .1).

Finally, our VLA images reveal that WNB1851+570 is in reality composed by two distinct fossil radio
galaxies. For both source we measured a spectral index $\alpha=2.3$. This association is a really
intriguing fact considering the rarity of this kind of sources.

\begin{figure*}
  \begin{center}

\includegraphics[width=16.5cm]{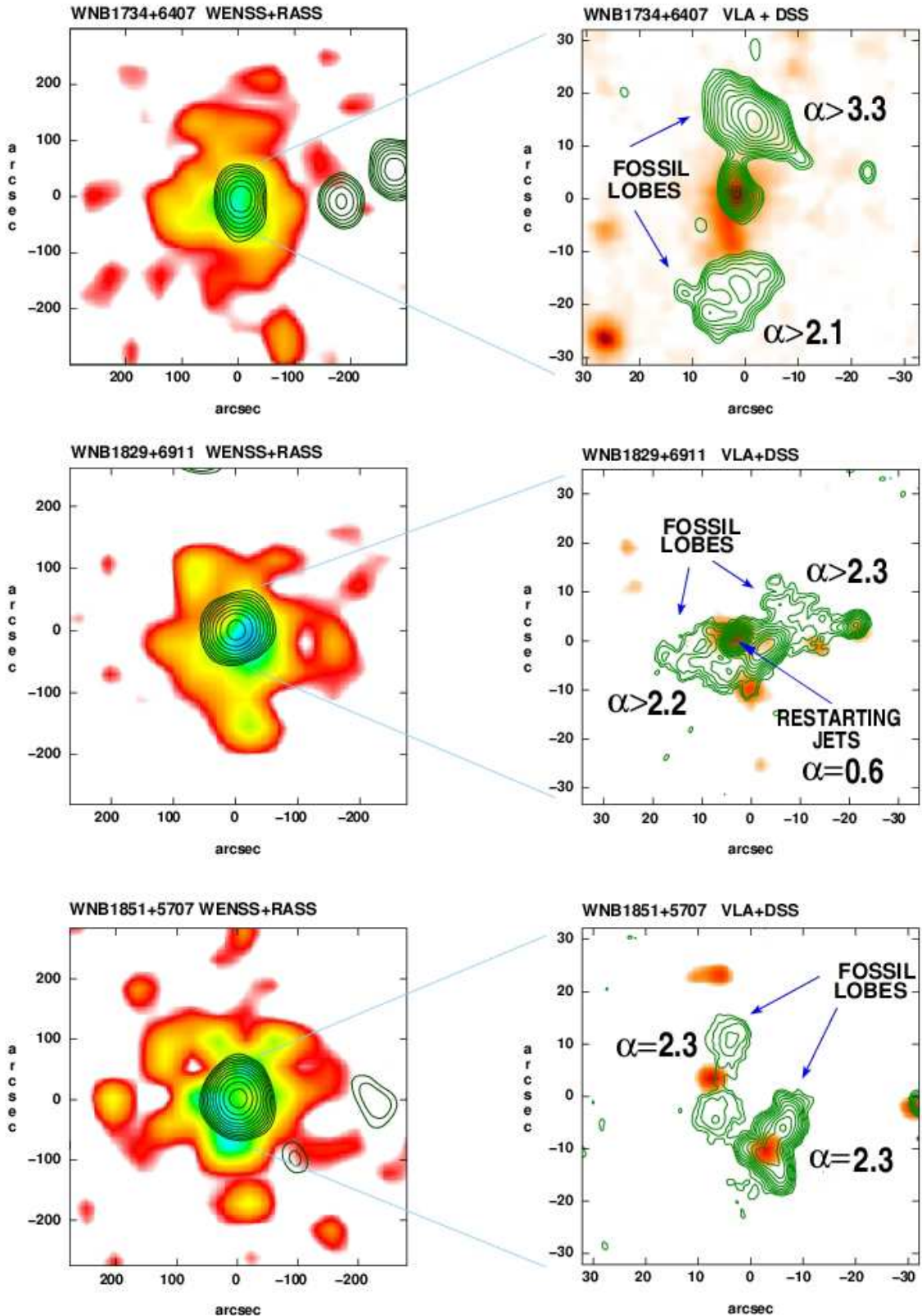}
    \caption{\small  Left panels: overlays of the WENSS (contours) and RASS (gray-scale). Right panels:
 overlays of the VLA 1.4-GHz contours with the optical images taken from the Digitized Sky Survey.
  The only plausible explanation for the extremely steep spectrum of the fossil lobes is that the
injection of relativistic electrons in these components is ceased for a significant fraction of the
source lifetime (see text).}
    \label{fig2}
  \end{center}
\end{figure*}

Some physical parameters\footnote{We adopted a cosmology
with $H_{\rm 0}$ =71 km s$^{-1}$Mpc$^{-1}$, $\Omega_{\rm \Lambda}=0.73$ and $\Omega_{\rm m}=0.27$.} of the sources are listed in Tab. 1.
The calculation of the source power at 151 MHz, where the energy losses of the synchrotron electrons are less dramatic,
place them near the FRI-FRII division (Fanaroff \& Riley 1974).

\section{The X-ray environment}
We have also made a search in the Rosat All-Sky Survey (RASS) for possible
X-ray counterparts. Each source is located, in  projection, at the
center of an X-ray emitting cluster (see Tab. 1 and Fig.~2 left column panels).

The gaseous environment in which radio galaxies are embedded affects their
morphology and regulates their expansion.  These effects are particularly
important in clusters of galaxies, where radio galaxies and the
X-ray-emitting intracluster medium (ICM) influence each other profoundly.
The ICM confines and distorts radio lobes and causes the jet bending. At the
same time, high resolution {\it Chandra} images have shown that the lobes of
central radio galaxies inflate cavities in the ICM of many cooling flow
clusters (e.g., McNamara et al.\ 2000, 2001; Blanton 2001; Heinz et al.\
2002, Johnstone et al.\ 2002), which is currently the favored mechanism for
the suppression of cooling flows (e.g., Churazov et al.\ 2000; David et al.\
2001; B{\"o}hringer et al.\ 2002).
Here we point out that interactions between the radio galaxies and the ICM
may have drastic and interesting consequences at the final stages of
a radio source's life. The simplest explanation for the association of our dying radio galaxies with
cluster is that the pressure of the dense ICM, perhaps a ``cooling flow'', prevents a quick liquidation
of the radio source through adiabatic expansion.
If so, it is likely that the signs of the interaction between the radio plasma and the surrounding gas
have survived in the form of X-ray cavities in the ICM. Our fossil sources might be the
progenitors of `ghost' bubbles, the buoyantly rising X-ray cavities that
are not associated with the central radio sources, such as those recently
discovered by McNamara et al.\ (2001) in A2597.

In order to test this hypothesis, we have proposed {\it Chandra} observations to compare the
actual fading radio structures with the X-ray image of similar angular resolution.

\section{Statistical analysis}
The space density of powerful radio galaxies deduced from a complete sample can be used to derive stringent upper limits to
radio source lifetimes (e.g. Schmidt 1966). In the WENSS minisurvey sample we count 90 radio galaxies with red magnitudes
 brighter than $m_{\rm r}<16$ with a flux density at 327 MHz greater than 30 mJy.
Of them, $N_{\rm clr}$\,=\,3 sources are found inside bright X-ray emitting clusters of galaxies in the list of B\"oringer 
et al. 2000. Two of them are dying sources.  Taking into account the volume occupied by the radio sources in cluster of galaxies and by the dying sources we find a density ratio of 0.6. This means that the relic 
phase {\it in clusters} lasts for about 60\% of the active phase duration. We note that this estimate is in good agreement with the radiative ages deduced from the integrated
  spectra (Sect. 2).
We repeated this calculation by considering the radio galaxies in the B2 sample that lie outside rich Abell galaxy clusters.
 In this case the density ratio of the volume occupied by the dying sources to the volume occupied by the total number of
 radio sources outside galaxy clusters is 0.06.
The relic phase duration for a radio galaxy in the field results to be only 6\% of the source lifetime.
These statistical considerations indicate that the probability to observe a dying radio galaxy in clusters is therefore increased by a factor of about  ten.

\section*{Acknowledgments}
The National Radio Astronomy Observatory is operated by Associated
Universities, Inc., under contract with National Science Foundation.

\end{document}